\documentstyle[multicol,aps,prb,epsf,graphics,psfig]{revtex}
\input epsf

\newcommand{\be}{\begin{equation}}
\newcommand{\bel}[1]{\begin{equation}\label{#1}}
\newcommand{\ee}{\end{equation}}
\newcommand{\bea}{\begin{eqnarray}}
\newcommand{\ba}{\begin{array}}
\newcommand{\eea}{\end{eqnarray}}
\newcommand{\ea}{\end{array}}

\begin{document}
\draft


\title{ Optimised Traffic Flow at a Single Intersection: Traffic
Responsive signalisation }
\author{M. Ebrahim Fouladvand, Zeinab Sadjadi and M. Reza Shaebani }

\address{ Department of Physics, Zanjan University, P.O.
Box 313, Zanjan, Iran.}

\date{\today}

\maketitle

\begin{abstract}

We propose a stochastic model for the intersection of two urban
streets. The vehicular traffic at the intersection is controlled
by a set of traffic lights which can be operated subject to
fix-time as well as traffic adaptive schemes. Vehicular dynamics
is simulated within the framework of the probabilistic cellular
automata and the delay experienced by the traffic at each
individual street is evaluated for specified time intervals.
Minimising the total delay of both streets gives rise to the
optimum signalisation of traffic lights. We propose some traffic
responsive signalisation algorithms which are based on the concept
of cut-off queue length and cut-off density.

\end{abstract}
\pacs{PACS numbers: 05.40.+j, 82.20.Mj, 02.50.Ga}

\begin{multicols}{2}
\section{Introduction}

Modelled as a system of interacting particles driven far from
equilibrium, {\it vehicular traffic } provides the possibility to
study various aspects of truly non-equilibrium systems which are
of current interest in statistical physics
\cite{css99,helbbook,kerner,tgf95,tgf97,tgf99}. For almost half
century, physicists have been challenged to understand the
fundamental principles governing the vehicular flow
\cite{tgf95,tgf97,tgf99,prigogine}. Recently, discrete models such
as {\it cellular automata } have provided a significant
theoretical framework for the discipline of traffic flow
modelling. The first cellular automata, the so-called BML model,
was introduced by {\it Biham, Middleton } and {\it Levine}. This
model described a simplified network of urban intersections
\cite{bml}. soon after, cellular automata, found their way in
highway traffic through the pioneering work of {\it Nagel} and
{\it Schreckenberg} \cite{ns92} which became the ancestor of many
papers in the literature ( for a review see ref.
\cite{css99,helbbook} and the references therein ). The BML model
itself was later generalized to take into account several
realistic features such as faulty traffic lights, independent
turning of vehicles, and green-wave synchronization
\cite{chung,tadaki,cuesta,nagatani,torok,horiguchi,freund,chopard}.
The Nagel-Schreckenberg and BML model were recently combined to
cast an upgraded version of city traffic models \cite{cs}. In a
very recent paper, the model is now extended to account for
different types of global signalisation
\cite{brockfeld}.\\
Despite the aforesaid efforts and those carried out by traffic
engineers, the subject of optimal signalisation of realistic urban
networks has not yet been comprehensively reviewed. In the above
approaches, the main concern has been focused on the global
strategies of the traffic network and frequently the role of
isolated intersection have been suppressed. We believe that the
optimisation of traffic flow at a single intersection is a
substantial ingredient towards a global optimisation. Isolated
intersections are fundamental operating units of the sophisticated
and correlated urban network and thorough analysis of them would
be advantageous toward the ultimate task of the global
optimisation of the city network. In this regard, our objective in
this paper is to analyze the traffic state of an isolated
intersection in order to find a better insight into the problem.
In addition to theoretical viewpoint, an investigation of isolated
intersection could be of practical importance. To a very good
approximation, marginal intersections in cities are unaffected by
other intersections and can be regarded as isolated ones.
Generally there are two basic types of control for traffic lights
at intersections: {\it fixed-cycle} and {\it traffic-responsive}
\cite{book,robertson}. Both of these methods can be implemented
via centralized or decentralized strategies. The application of
each method strongly depends on traffic condition and the topology
of the city network. In this paper we study the basic features of
traffic flow at a single intersection which is controlled both
under fixed-cycle as well as traffic responsive schemes.

\section{ Formulation of the Model}

An isolated intersection is formed at the intersection of two
streets. The streets, in principle, can each carry two opposite
flows of vehicles. Depending on the designing of the intersection,
different phases of movement can be defined ( a phase of traffic
is defined as the flow of vehicles that proceed an intersection
without conflict). Here for simplicity we restrict ourselves to
the simplest structure: a {\it one-way to one-way } intersection.
With no loss of generality, we take the direction of the flow in
the first street, hereafter referred to as the street A,
northwards.  The other street ( hereafter referred to as street B)
conducts a one-way eastward flow. Cars arrive at the south and
west entrances of the intersection. Space and time are discretized
in such a way that each chain is divided into cells which are the
same size as a typical car length. Time is assumed to elapse in
discrete steps. We take the number of cells to be $L$ for both
roads. Each cell can be either occupied by a car or being empty.
Moreover, each car can take discrete-valued velocities
$1,2,\cdots, v_{max}$. To be more specific, at each step of time,
the system is characterized by the position and velocity
configurations of cars and the traffic light state at each road.

\begin{figure}\label{Fig1} \epsfxsize=7.5truecm
\centerline{\epsfbox{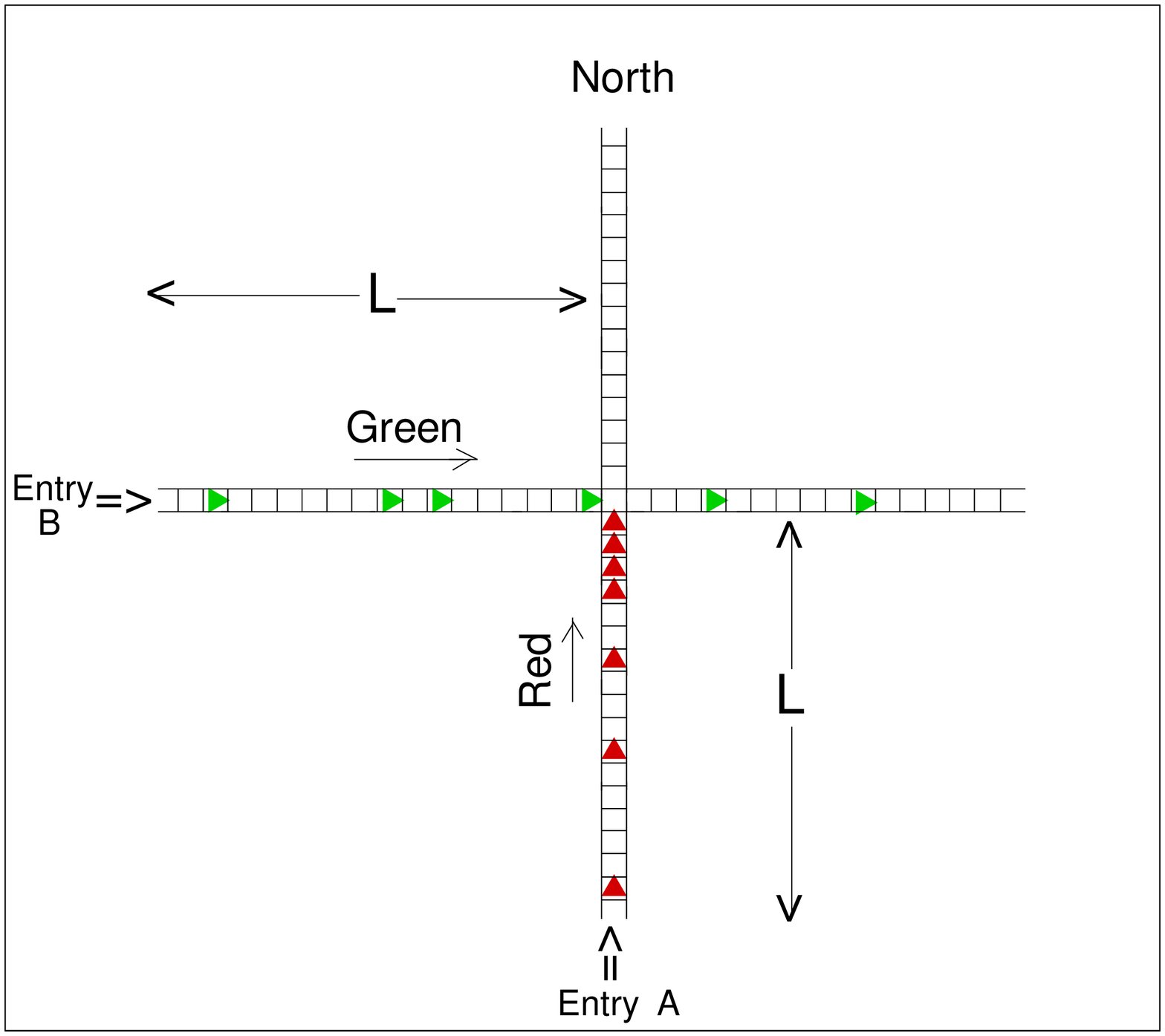}}
\end{figure}

The system evolves under a generalized discrete-time
Nagel-Schreckenberg (NS) dynamics. The generalized model
incorporates anticipation effects of driving habits\cite{knospe}.
It modifies the standard NS model at its second step i.e.,
adjusting the velocities according to the space gap. Let us
briefly explain the updating rules which are synchronously applied
to all the vehicles. We denote the position, velocity and space
gap (distance to its leading car) of a typical car at discrete
time t by $x^{(t)},v^{(t)}$ and $g^{(t)}$ and the same quantities
for its leading car by $x_l^{(t)},v_l^{(t)}$ and $g_l^{(t)}$.
Assuming that the expected velocity of the leading car,
anticipated by its follower, in the next time step $t+1$ takes the
form $v_{l,anti}^{(t)}=min(g_l^{(t)},v_l^{(t)})$, we define the
effective gap as $g_{eff}^{(t)}:= g^{(t)} +
max(v_{l,anti}^{(t)}-gap_{secure},0)$ in which $gap_{secure}$ is
the minimal security gap. Concerning the above considerations, the
following updating steps evolves the position
and the velocity of each car.\\

1) Acceleration:\\

$v^{(t+1/3)}:= min(v^{(t)}+1, v_{max})$\\

2) Velocity adjustment :\\

$v^{(t+2/3)}:=min(g_{eff}^{(t+1/3)}, v^{(t+1/3)})$\\

3) Random breaking with probability $p$:\\

if random $< p$ then $v^{(t+1)}:=max(v^{(t+2/3)}-1,0)$\\

4) Movement : $x^{(t+1)} :=x^{(t)}+ v^{(t+1)}$ \\

Let us now specify the physical value of our time and space units.
Ignoring the possibility of existence of long vehicles such as
buses, trucks etc, the length of each cell is taken to be as 5.6
metres which is the typical bumper-to-bumper distance of cars in a
waiting queue. Concerning the fact that in most of urban areas a
speed-limit of 60 kilometre/hour should  be kept by the drivers,
we quantify the time step in such a way that $v_{max}=6$
corresponds to the speed-limit value (taken as 60 km/h). In this
regard, each time step equals two seconds and therefore each
discrete increments of velocity signifies a value of 10km/h which
is equivalent to a comfortable acceleration of 1.4 $m/s^2$. We
have also set the horizon length $L=70$ cells and $gap_{sec}=1$.
The state of the system at time $t+1$ is updated from that in time
$t$ by
applying the modified NS dynamical rules.\\

{\bf Step 1 : signal determination}.\\

We first specify the signal states for all of the driving
directions. In subsequent sections we will, in detail, explain the
scheme at which the traffic lights change their colour.\\

{\bf Step 2 : movement in the green road}.\\

At this stage, we update the position and velocities of cars on
the green road according to the movement rules which are
synchronously applied to each car .\\

{\bf Step 3 : movement at the red road, delay evaluation }.\\

Here the updating is divided into two parts. In the first part, we
evaluate the delay of cars waiting on the red period of the
signalisation. In the second half, we update the position and
velocities of the moving cars approaching the waiting queue. We
should note that once the signal switches to red, the moving cars
continue their movement until they come to a complete stop when
reaching to the end of the waiting queue. As soon as a car comes
to halt, it contributes to the total delay. In order to evaluate
the delay, we measure the queue length (the number of stopped
cars) at time step t and denote it by the variable $Q$. We recall
that $ pos^{red}[i,t]=1 $ for $i=1 \cdots Q $ and zero at $i=Q+1$.
Delay at time step $t+1$ is obtained by adding the queue length
$Q$ to the delay at time step $t$. \be delay(t+1)= delay(t)+ Q(t)
\ee

This ensures that during the next time step, all the stopped cars
contribute one step of time to the delay. The next part describes
the update of positions and velocities of moving cars. Moving cars
can potentially be found in the cells $ Q+2,Q+3, \cdots, L$. We
update their positions and velocities accordingly.\\

{\bf step 4 : entrance of cars to the intersection}.\\

So far, we have dealt with those cars within the horizon of the
intersection which goes up to the entry point located at site $L$.
Here we discuss the entrance of cars into the intersection.
Generally speaking, during each green period, a fraction of the
queue will dissolve and pass the intersection. If the average
arrival rate of cars exceeds the maximum evacuation rate of the
lane, then, on an average level, a fraction of a generic queue
will not be able to go through the intersection and should wait
until the next upcoming green period arrives. Correspondingly in
the course of time, the remainders accumulate giving rise to a
growing queue length. In reality we rarely observe this phenomena
since an actual intersection is linked to other ones hence the
possibility of such a rare event is restricted to very exceptional
cases where there is an overwhelming large amount of incoming
flux. Throughout the paper, we assume that the average in-flow
rate is sufficiently below this evacuation rate so that a generic
queue will have enough green time to dissolve and that the
intersection is able to support the incoming flux. In this case,
there is a typical queue length which affects the motion of
incoming cars. The motion of a car approaching the red light is
affected via two factors: the distance to the end of the queue and
the traffic light signal which act as a hindrance to the incoming
traffic flow. In the light traffic state under consideration,
there exist an interaction distance to the queue beyond which one
can, to a good extent, assume that the cars are moving without
being affected by the traffic light signal. Away from this
interaction zone, the cars move according to the movement rules
without any hindrance. We take the position of the place at which
the cars enters the horizon of the intersection to be 70 cell
equivalent to 400 metres. The time head-ways between entering cars
at this entry location vary in a random manner which consequently
implies a random distance headway between successive entering
cars. As a candidate for describing the statistical behaviour of
random space gap of entering cars, we have chosen Poisson
distribution. The Poisson distribution function has been used in a
variety of phenomena incorporating the modelling of "queue
theories" and has proven to be a good estimation of reality
\cite{erlang}. In addition, it has the merit of taking only
discrete values which is desirable to us in the view of the fact
that in our model the gap is a discrete variable. According to
this distribution function the probability that the space gap
between the car entering the intersection horizon and its
predecessor be $n$ is : $ p(n) = \frac{\lambda^n e^{-\lambda}
}{n!} $ where the parameter $\lambda$ specifies the average as
well as the variance of distribution function. The parameter
$\lambda$ is a direct measurement of traffic volume. A large value
of $\lambda$ describes a light traffic while on the other hand, a
small-valued $\lambda$ corresponds to a heavy traffic state. Now
we let the cars enter into the intersection's horizon. At the end
of the movement rule, we evaluate the position of farthest cars on
both streets. We denote them by $last_A$ and $last_B$. By
definition , $pos_A[j,t]=0$ for $j>last_A$.  A similar statement
applies to street B. In order to simulate the entrance of cars
into the horizon of the intersection, we randomly choose an
integer weighted by Poisson distribution function. This number
represents the gap of the oncoming successor of the farthest car.
Let us denote these numbers (headways) by $h_A$ and $h_B$ for
street A and B respectively. Once the random gap is chosen, we
create a car at the position $last_A+h_A$ ($last_B+h_B$) of the
street A (B) respectively. The created car survives provided the
following constraint is satisfied: $ last_A +h_A \leq L ~~$ (~~$
last_B+h_B\leq L $). If the position of the created car exceeds
the horizon length $L$, the creation procedure is rejected. The
above {\it ad hoc} rules updates the configuration of the
intersection in time. In the next section, we will explain our
simulation results.

\section{ Signalisation of traffic lights: Fixed Time Scheme }

In this controlling scheme, the traffic flow is controlled by a
set of traffic lights which are operated in a fixed-cycle manner.
The lights periodically go green with a fixed period (cycle
length) $T$. This period is divided into two parts : in the first
part, the traffic light is green for street $A$ (simultaneously
red for street $B$). This part lasts for $T_{g}$ seconds ( $ T_{g}
< T $). In the second part, the lights change colour and the
movement is allowed for the cars of road B. The second part lasts
from $T_{g}$ to $T$. This behaviour is repeated periodically. Cars
enter the intersection and a fraction of them experience the red
light and consequently have to wait until they are allowed to go
through intersection during the upcoming green
periods. Now we raise the basic question: \\

{\it how should one adjust the ratio of $\frac{T_{green}}{T}$ in
order to optimise the
throughput flow?}.\\

There is now an almost well-established agreement on the
quantitative definition of optimisation. Borrowing from the
traffic engineering literature, we adopt {\it optimised traffic }
as a state in which the total delay of vehicles is minimum. In one
of our preceding works \cite{foolad}, we analytically evaluated
the total delay in terms of arrival as well as that of the exit
rates of vehicles. However, our approached was based on the simple
assumption of the time-constancy of the arrival rates. In reality,
we know that successive cars arrive with fluctuating time-headways
which consequently induces time-varying arrival rates. In this
paper we address the question of non-constant rates. In order to
evaluate the delay, we have simulated the flow of vehicles. For
the sake of simplicity, we have assumed that each street has a
single lane. For streets with more than one lane, one simply
should multiply the value of delay by the number of lanes.

\subsection{ Simulation Results: Symmetric inflow }

We let the intersection evolve for 1800 time steps which is equal
to a real time period of one hour. We let the green time of street
A, $T_{g}$, vary from zero to $T$. For each value of $T_{g}$, we
evaluate the aggregate delay corresponding to both traffic lights
during 1800 time steps. We have averaged the results of 50
independent run of the programme. Let us first consider the
symmetric traffic states in which the traffic conditions are equal
for both roads. In this case, we equally load the intersection
with entering cars spatially separated by random space gap
(Poisson statistics) from each other. The following graphs depicts
the total delay curves as a function of $T_{g}$
allocated to road A for two values of cycle length.\\

\begin{figure}\label{Fig2}
\epsfxsize=7.5truecm \centerline{\epsfbox{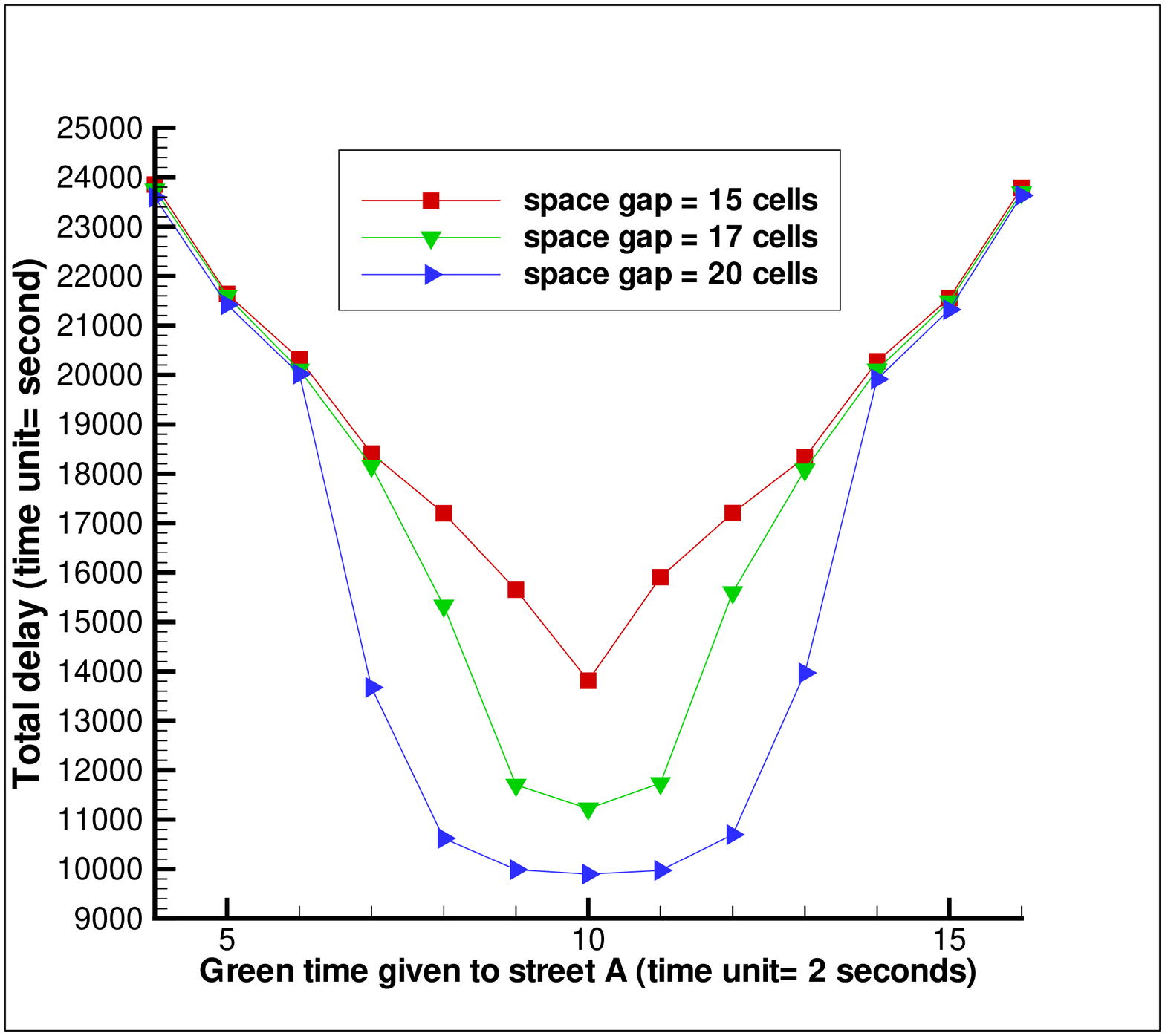}}
\end{figure}

\begin{figure}\label{Fig3}
\epsfxsize=7.5truecm \centerline{\epsfbox{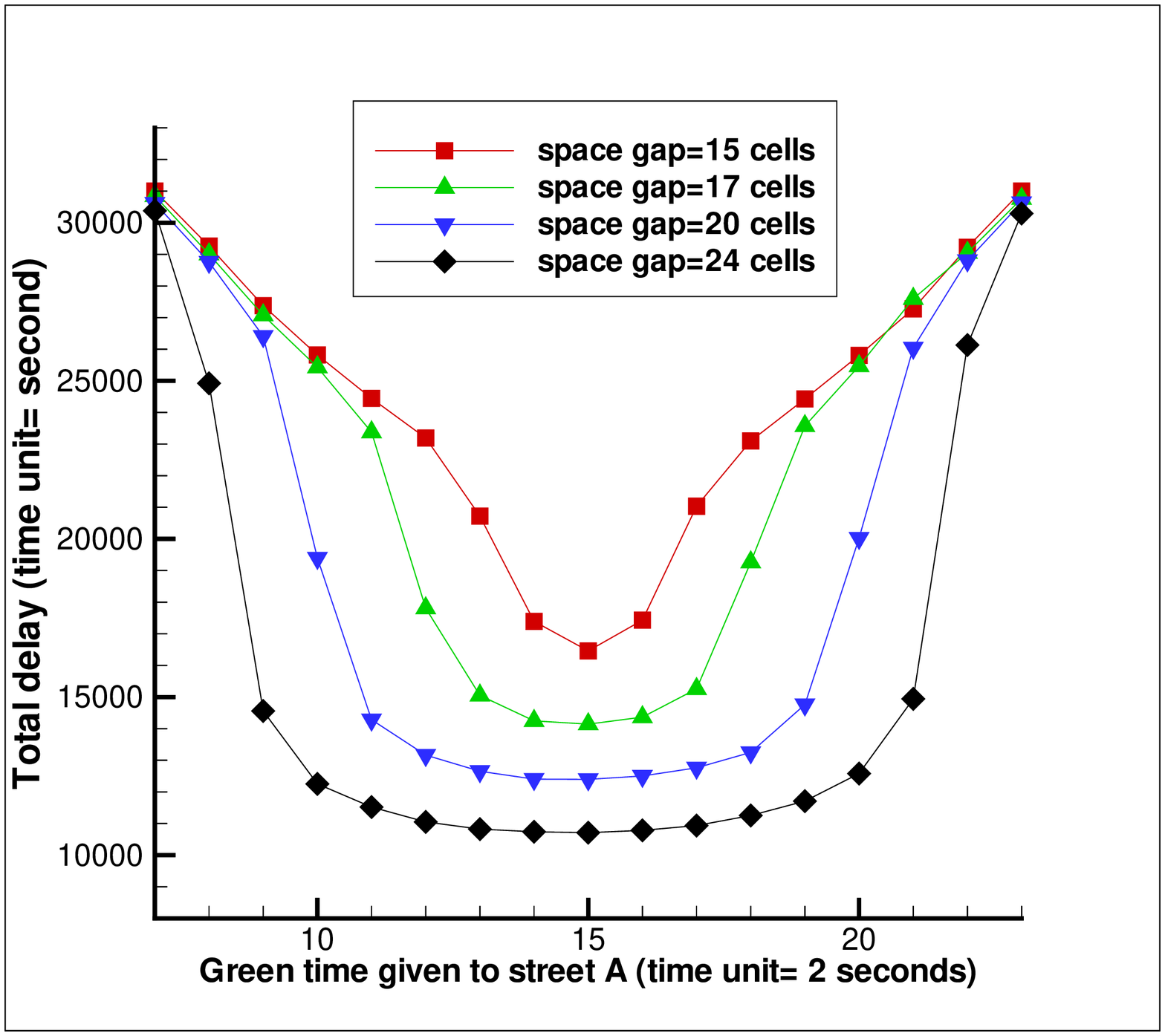}}
\vspace{0.02cm} {\small{Fig.~2 and 3: The cycle length is 40
seconds (top graph) and 60 seconds (bottom graph). Total delay
versus the green time of road A is sketched for various in-flow
rates.} }\\
\end{figure}

The general shape of the total delay curve resembles a "U-shaped"
valley. The middle area, where the total waiting time is
minimised, corresponds to a situation in which the evacuation rate
of roads exceeds the in-flow and the queues can be dissolved
during one green period and consequently all the waiting cars will
be able go through the intersection in the upcoming green period.
The optimal traffic flow is obtained by keeping the lights at
equal timing $T_g=\frac{T}{2}$. Before preceding further, it would
be useful to discuss, in detail, the conditions in which the
intersection would not able to support the in-flux and hence the
queue length starts growing. Supposing the fraction
$\frac{T_g}{T}$ of the cycle is given as green time to a street
say A. Therefore during the one hour time interval, street A
receives $3600\frac{T_g}{T}$ seconds of green times. Using the
fact that in green phase vehicles are going through the crossing
with the approximate rate of 0.5/sec/lane, on an average level,
the maximum out-flow capacity, denoted by $ \langle C_{max}\rangle
$ would be: $\langle C_{max}\rangle = 1800 \frac{T_g}{T}$. On the
other hand, total number of in-flow can be estimated once the
average space gap is given. To do so, we first obtain the average
time headway of entering vehicle (in second), denoted by
$\Delta_{\lambda_A}$, as follows $ \Delta_{\lambda_A}= 5.6
\frac{\lambda_A}{v_{max}}$. In the above formula, we have assumed
that entering cars have the maximum velocity and 5.6 denotes the
cell length in metre. Concerning the above consideration, during
one hour, the average number of entering cars are approximated by
$\langle N_{in}\rangle=\frac{3600}{\Delta_{\lambda_A}}$. Avoiding
the occurrence of growing queue, leads to the satisfaction of the
following constraint \cite{foolad}:

\be \langle N_{in}\rangle \leq \langle C_{max} \rangle \ee

Substituting maximum velocity by 16.7 m/s in the above constraint,
gives rise to condition $ T_g \geq \frac{6T}{\lambda_A} $. Taking
$T=60$ seconds, the minimum consistent value of $T_g$ for $
\lambda=13$, 16 and 19 cells would be 27, 22 and 19 seconds
respectively. Alternatively, the lowest $\lambda_A$ is restricted
to 12 cells or 67 metres. Similar arguments should be applied to
street B. Concerning the fact that the green time of street B is
$T-T_g$, one simply deduce that the condition $ T- T_g \geq
\frac{6T}{\lambda_B}$ should hold. Combining the two conditions on
$T_g$, one arrives at the following inequality on $T_g$ for a
stationary condition of the queues.

\be \frac{6T}{\lambda_A} \leq T_g \leq T(1-\frac{6}{\lambda_B})
\ee

Consequently, the allowed $\lambda_A$, $\lambda_B$ and $T_g$
should satisfy the above inequality, otherwise the queue will grow
infinitely in time. From the above relation one concludes two
necessary conditions on the rates: $\lambda_B \geq 6$ and
$\frac{1}{\lambda_A} + \frac{1}{\lambda_B} \leq \frac{1}{6}$ which
hold regardless of the value of green time. We note that the above
arguments are based on simple mean-field approximation and no
fluctuation has been taken into account. There are two origins of
fluctuation, the first one concerns the stochastic space gap of
cars which make the incoming flux deviated from a constant one,
and second, the interaction among cars and randomness arising from
the car movements. Therefore the region close to critical volume
$\lambda_A=12$ needs special treatment. Simulation results shows
that away from the critical region, fluctuations do not violate
our mean-field conclusions.

\subsection{ Asymmetric in-flow rates }

In the asymmetric states, the entering rates into the roads differ
from each other. An interesting asymmetric state is the
intersection of a major to a minor street. A large fraction of
urban intersections lies in the category of major-to-minor. The
signalisation of these types of crossing is still a controversial
subject. The main reason is that in most of intelligent real-time
controller systems, the signalisation of these intersections are
highly affected by the major intersections signalisation schemes
which frequently overlook the local optimisation of minor
intersections. In our model, a major-to-minor crossing is modelled
by a light traffic in one road and a congested one in the other
road. The following graph depicts the behaviour of the delay curve
at a major-to-minor intersection. As observed, the minima of delay
curves are shifted toward right which expresses the fact that the
majority of green times should be given to the road with higher
in-flow rate. Analytical considerations lead to distribution of
green times in proportion to the in-flow rates of the roads
\cite{foolad}. To be more specific, let us denote the average
arrival rate at street A and B by $\alpha_A$ and $\alpha_B$. Note
that $\alpha_A$ and $\alpha_B$ are proportional to inverse of
entering space gaps. We now evaluate the waiting time of street A
during one complete cycle. The number of cars arriving the queue
during the time interval $[t,t+dt]$ in the red period, which lasts
$T-T_g$ seconds, is $\alpha_A dt$. These cars should wait
$T-T_g-t$ seconds until the upcoming green period starts.
Therefore, their contribution to delay reads $\alpha_A
dt(T-T_g-t)$. The aggregate delay of street A during the cycle is
obtained by summation of infinitesimal delays as follows:\\

\be \int_0^{T-T_g} \alpha_A dt(T-T_g-t) =
\frac{1}{2}\alpha_A(T-T_g)^2 \ee

Similarly the contribution of street B to total delay can simply
be evaluated as $\frac{1}{2}\alpha_BT_g^2$. Adding the street
delays together and minimising it with respect to $ T_g$ gives
rise to the optimal $T_{g}$:\\

\be T_g^{opt} = \frac{\alpha_A}{\alpha_A+ \alpha_B} T \ee

The above result corresponds to distribution of green times in the
ratio of in-flow rates. Our simulations results are in good
agreement to this conclusion. In fact the position of optimal
green times, i.e., the minima of the delay curves, grow linearly
with the linear decrease of in-flow rates which are in turn
proportional to the inverse of average space gap.

\begin{figure}\label{Fig4}
\epsfxsize=7.5truecm \centerline{\epsfbox{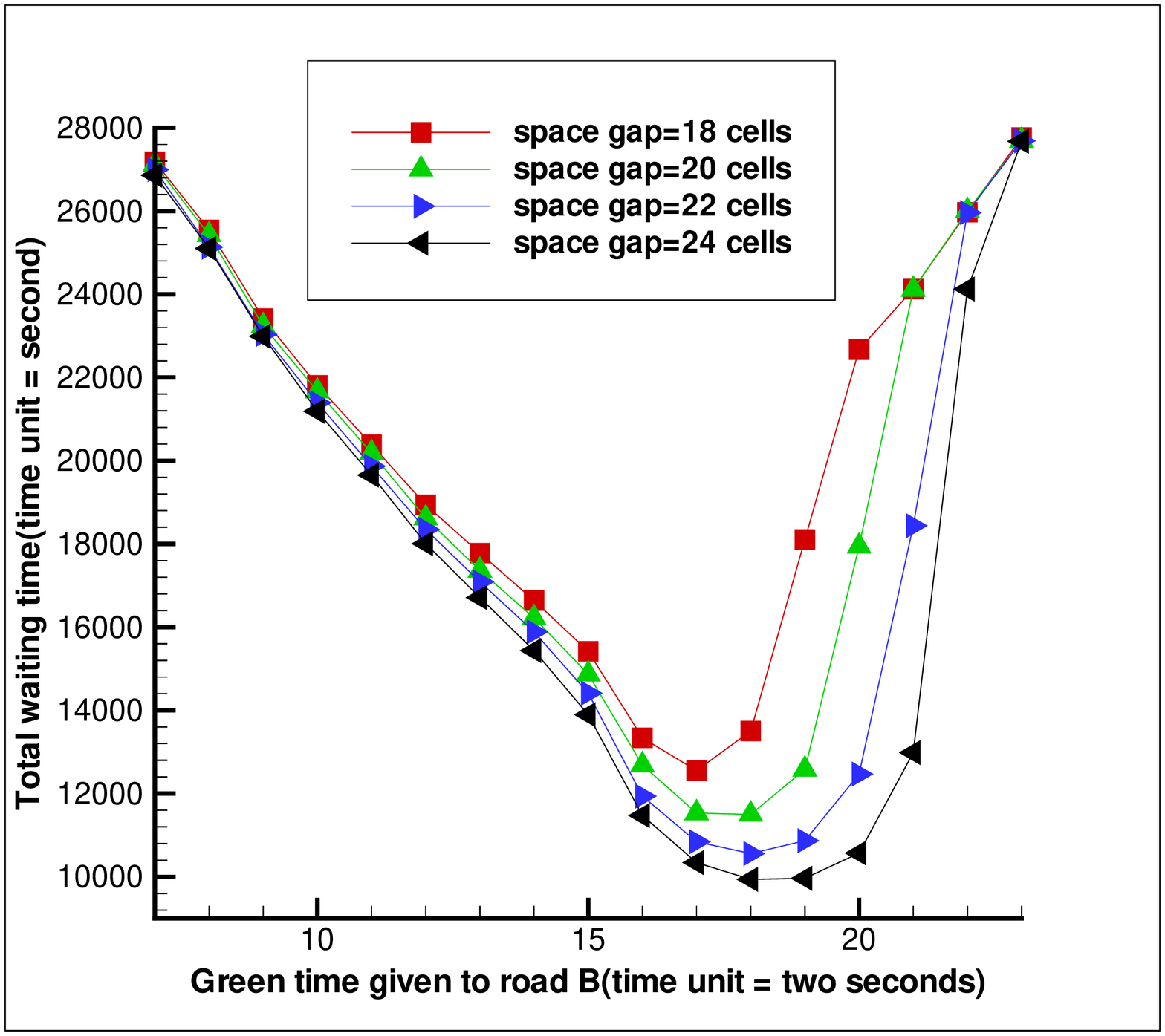}}
\end{figure}
\vspace{0.02 cm} {\small{Fig.~4: Total delay for asymmetric
traffic volumes in a major-to-minor crossroads. The space gap of
the major road is set to 13. The space gap of the minor road is varied. T=60 seconds.} }\\

\section{ Signalisation of traffic lights: traffic adaptive }

We now discuss the traffic adaptive controlling scheme in which
the light signalisation is adapted to the traffic at the
intersection. Nowadays advanced traffic control systems anticipate
the traffic approaching intersections. Traffic-responsive methods
have shown a very good performance in controlling the city traffic
and now a variety of schemes exists in the literature
\cite{book,robertson,porche,huberman}. In these schemes, the data
obtain via traffic detectors installed at the intersection is
gathered for each movement direction and it is possible to count
the queue lengths formed behind the red lights. One can also
measure the time-headways between successive cars passing each
lane detector. Thus it is possible to estimate the traffic volume
existing at the intersection. There are various methods for
distribution of green times. In what follows we try to explain
some standard ones. In each scheme, the green time of a typical
green street is terminated if some conditions are fulfilled. By
green (red) street we mean the street for which the traffic light
is green (red). We now state the termination conditions for each scheme.\\

{\bf Scheme (1)}: The queue length in the conflicting direction
exceeds a cut-off value $L_c$. This scheme only concerns
the traffic states in the red street.\\

{\bf Scheme (2)}: The global car density in the green street falls
below the cut-off value $\rho_c$. Here the algorithm
only considers the traffic state in the green street.\\

{\bf Scheme (3)} : Each direction is endowed with two control
parameters $L_c$ and $\rho_c$. The green phase is terminated if
the conditions: $ \rho ^{g} \leq \rho_c $ and $L^{r} \geq L_c$ are
both satisfied.\\

Here the algorithm implements the traffic states in both streets.
The superscripts "r" and "g" refer to words "red" and "green"
respectively. We note that the first two schemes are special cases
of the more general scheme (3). Schemes (1) and (2) are the
limiting cases of schemes (3) by letting $\rho_c \rightarrow 1$
and $L_c \rightarrow 0$ respectively. In general, the numerical
value of control parameters $L_c$ and $\rho_c$ could be taken
different for each individual street. In what follows we present
our simulation results for different types of signalisation
schemes introduces above.

\subsection{ Simulation results: symmetric inflow }

We let the intersection evolve for 1800 time steps which is equal
to a real time period of one hour. We evaluate the aggregate delay
for both streets as well the number of passed vehicle during 1800
time steps at each direction for different traffic situations. As
the first situation, we have considered the symmetric traffic
state in which the traffic conditions are equal for two streets.
In this case, we equally load the streets with entering cars
spatially separated by random space gap, obeying Poisson
statistics, from each other. We first discuss the first scheme in
which the lights change colour as soon as the queue length reaches
the cut-off length. We set an equal cut-off length for both
streets. The following graph depicts the total delay curve as a
function of traffic volume for various cut-off lengths.\\

\begin{figure}\label{Fig5}
\epsfxsize=7.5truecm \centerline{\epsfbox{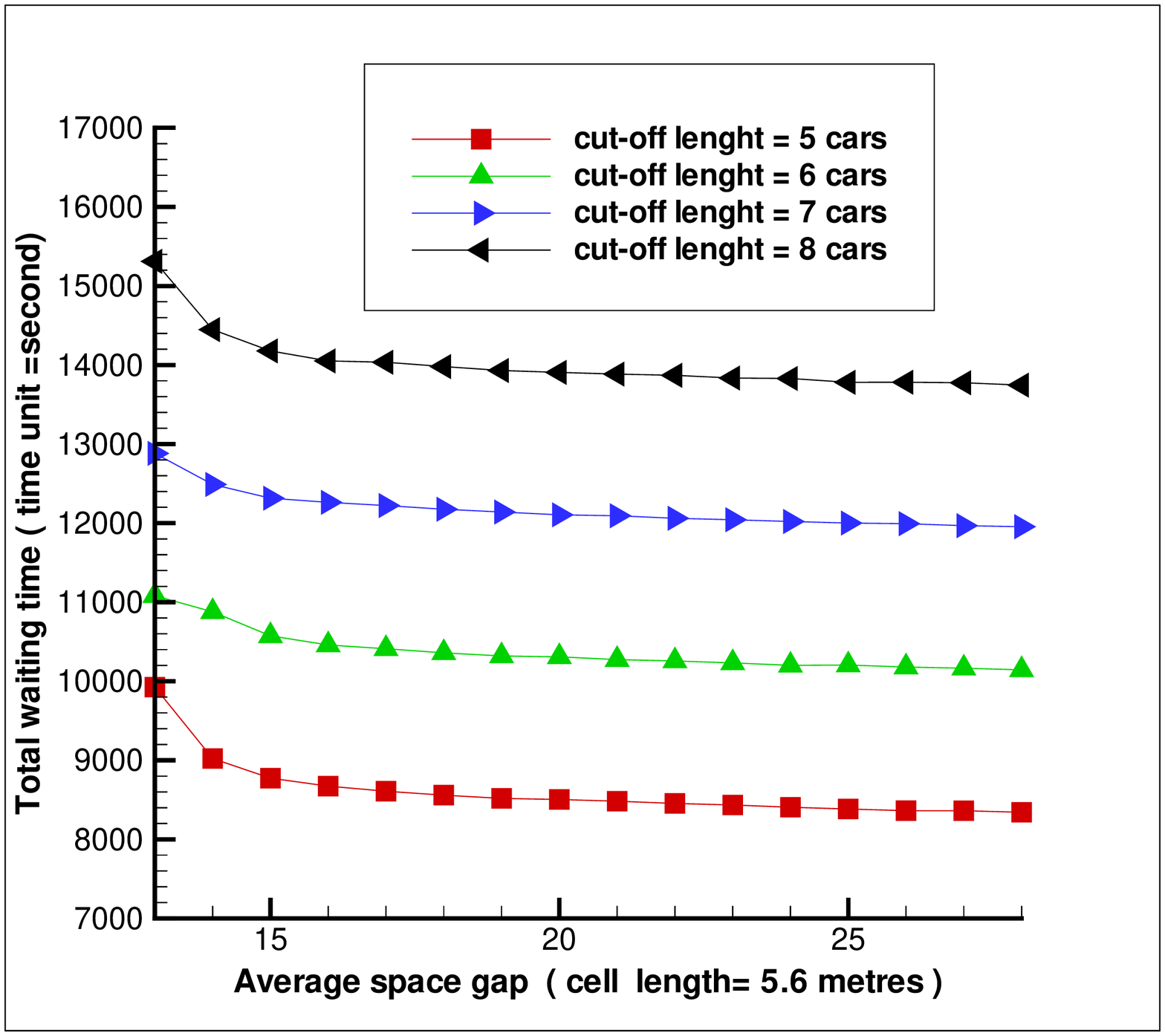}}

\vspace{0.02 cm} {\small{Fig.~5:  total delay in terms of average
space gap of entering cars for some cut-off queue length $L_c=
5,6,7,8$. }}
\end{figure}

We observe that for each cut-off length, total delay shows a
slight decrease with respect to the in-flow rate. The graph
demonstrate that the shorter the cut-off length, the less the
total delay. The next graph shows the delay versus in-flow rate
when the lights are signalised according to scheme (2). Here the
green light is terminated below a certain occupancy in the
corresponding moving flow.

\begin{figure}\label{Fig6}
\epsfxsize=7.5truecm \centerline{\epsfbox{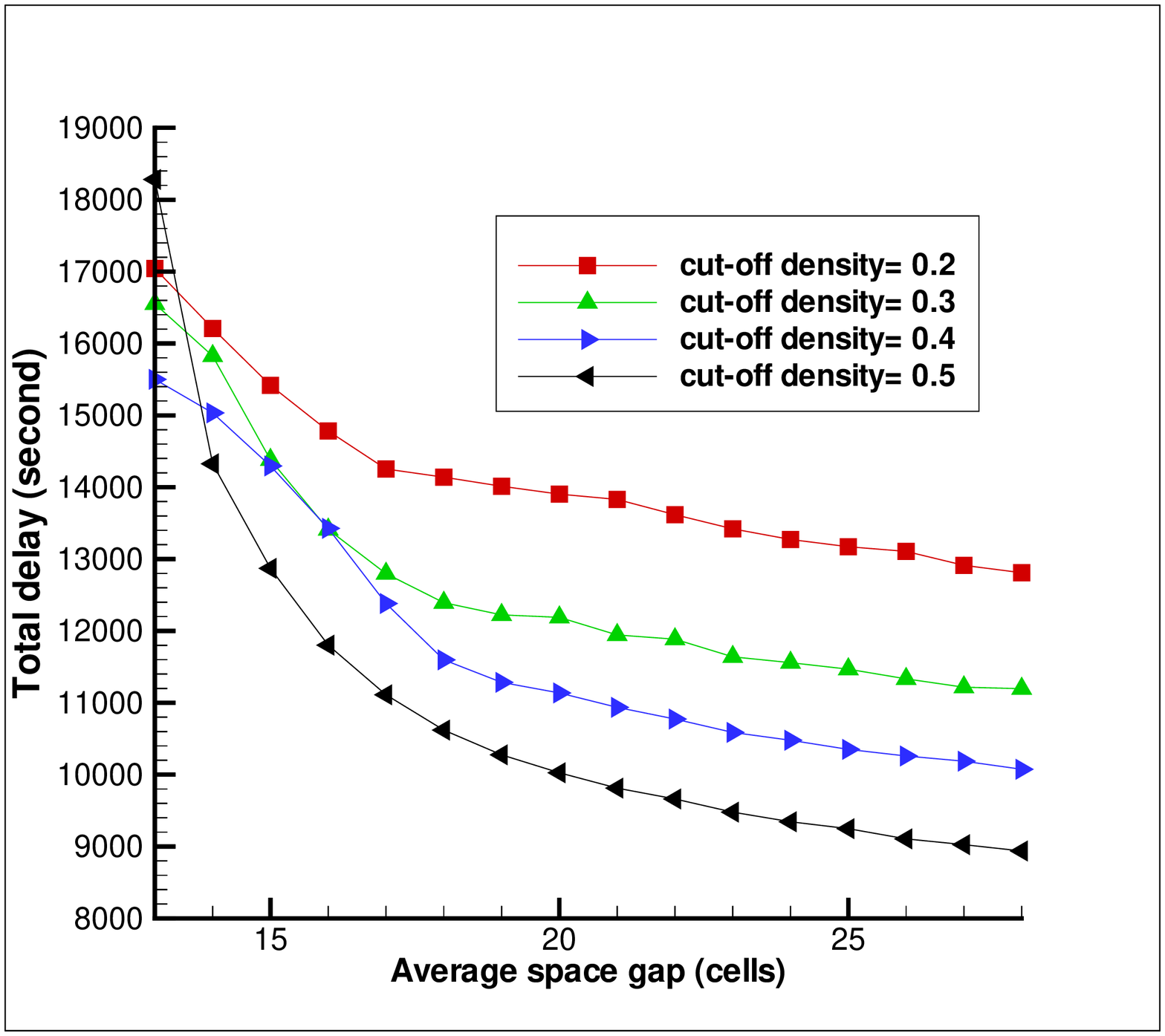}}

\vspace{0.02 cm} {\small{Fig.~6:  total delay in terms of average
space gap of entering cars for some cut-off densities. }}
\end{figure}

We observe that the first scheme gives rise to lower delay. The
reason is that in relatively low traffic volume, more cars have to
stop in the red light in order that the global density in the
green direction reaches to its cut-off value. This raises the
delay compared to scheme (1). Finally we discuss the third scheme
in which two conditions should be satisfied for termination of
green light. In this scheme both cut-off densities and cut-off
lengths are taking into account. We have investigated the case
where cut-off parameters are equal for both streets i.e.,
$\rho_c^{A}= \rho_c^{B}=\rho_c$ and $L_c^{(A)}=L_c^{(B)}=L_c$. The
following table shows the delay table of one-hour performance for
various
values of  $\rho_c$ and $L_c$. Traffic in-flow rates are $\lambda_A= \lambda_B=13$.\\

\begin{tabular}{|c|c|c|c|c|c|}
\hline
$\rho_c \downarrow ~ | ~ L_c \rightarrow $ & 5 & 6 & 7 & 8& 9  \\
\hline
0.10 & 14000 & 15500 & 15900 & 17900 & 18600  \\
\hline
0.20  & 13200 & 14900 & 14700 & 16500 & 17400 \\
\hline
0.30  & 11500 & 12800 & 13200 & 15500 & 16400 \\
\hline
0.40 & 11200 & 11900 & 12600 & 14000 & 15300  \\
\hline
0.50 & 10900 & 11200 & 11700 & 12700 & 13900\\
\hline
\end{tabular}
\vspace{.5cm}

The above result sets the optimal parameters at $\rho_c=0.5$ and
$L_c=5$. We have tested a variety of other traffic volumes. The
results demonstrate that in the control parameter space, the
optimal region lies in high $\rho_c$ and low $L_c$. The above
result shows that scheme (3) gives an improved delay with respect
to scheme (2). Our simulation results shows that this conclusion
can be made for other values of $\lambda$ hence the third scheme
is more optimal than the second one. However the results show that
still the first scheme gives lower delays compared scheme (3).
Specifically for $\lambda=20$ cells, the predicted delays are
8500, 10100 and 9600 seconds respectively. The next graphs depict
the behaviour of consecutive green times of streets. We have
chosen the green times of street A. Fluctuating traffic volume at
streets induces fluctuating green periods which are known as the
{\it green time plans} in the traffic engineering literature
\cite{robertson,huberman}. We note that a realistic optimising
algorithm must satisfy the constraint that the duration of green
phases should not be shorter that a minimal amount e.g. 10 sec or
so. This minimal value reflects the actual time required for an
immobile waiting queue to accelerate and make a considerable
movement forward. Our simulation results shows that $L_c \geq 4$
corresponds to typical green periods greater than 12 seconds. It
may be useful to analyze the statistics of green plans. Denoting
the consecutive green times by $T_g^{(1)}, T_g^{(2)},T_g^{(3)}
\cdots$, We first obtain the basic statistical properties which
are the moments of the distribution of the green times. The bottom
figure shows the green plan histogram of street A for $\lambda_A=
\lambda_B=13$. For the mentioned traffic state, the average green
time and the standard deviation of street A are 18 and 2.54
seconds respectively.

\begin{figure}\label{Fig7}
\epsfxsize=7.5truecm  \centerline {\epsfbox{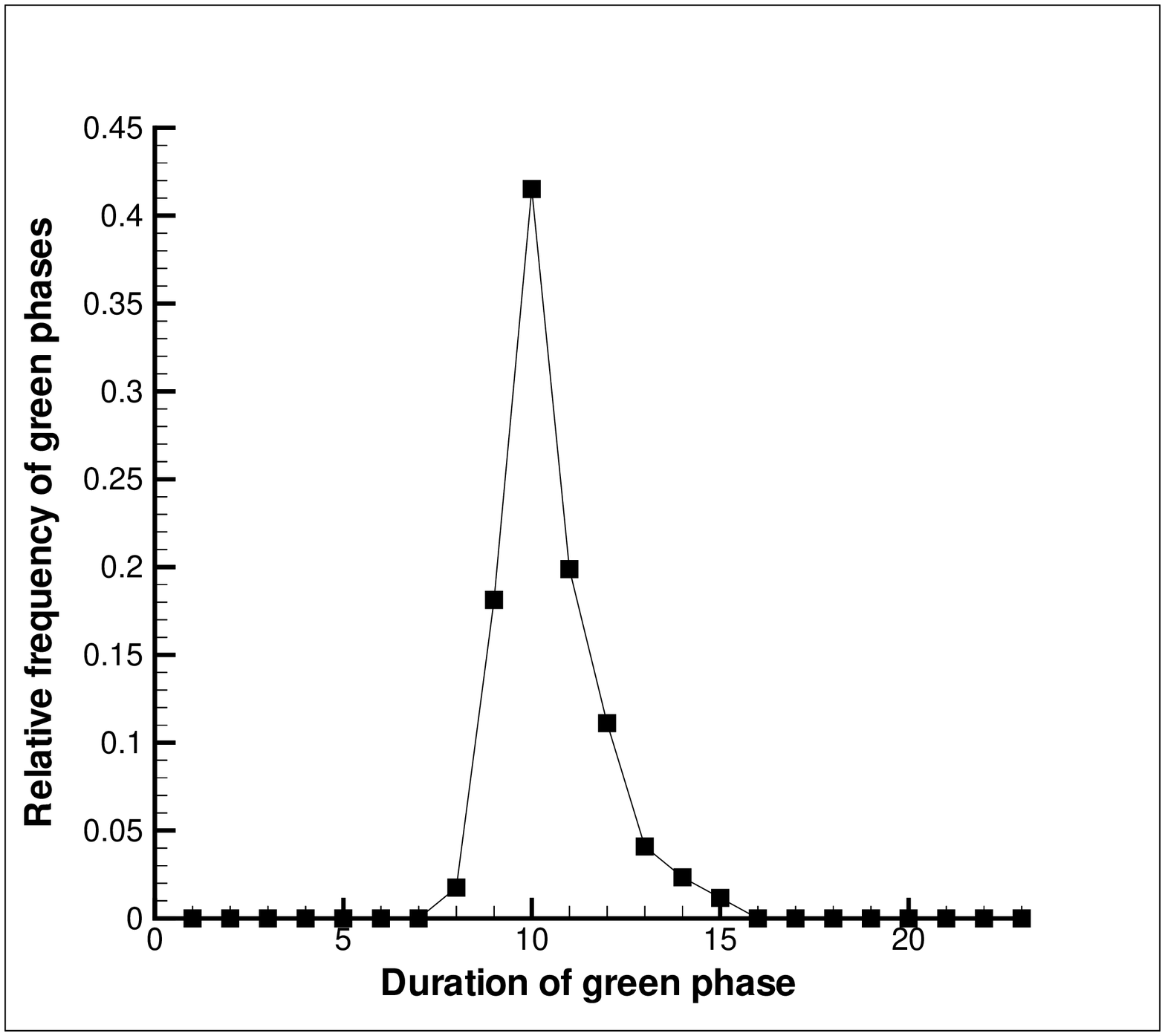}}

\end{figure}

\vspace{0.02 cm} {\small{Fig.~7: Green times histogram of street
A. $\lambda_A=\lambda_B=13$ cells. The lights are controlled via
scheme (1). The cut-off length is taken as 7 cars.}}

\subsection{Asymmetric in-flow}

{\bf scheme (1)}:\\

We now investigate the situations where the streets are not
equally loaded. For this purpose, we fix the in-flow rate of
street A at $\lambda_A= 13 $ and vary the in-flow rate of street
B. The following graph exhibits the total waiting time curves as
well as those of each street for some cut-off lengths (taken equal
for both streets). Here again the less delay is achieved for
shorter cut-off lengths which in turn correspond to shorter green
phases.

\begin{figure}\label{Fig8}
\epsfxsize=7.5truecm  \centerline {\epsfbox{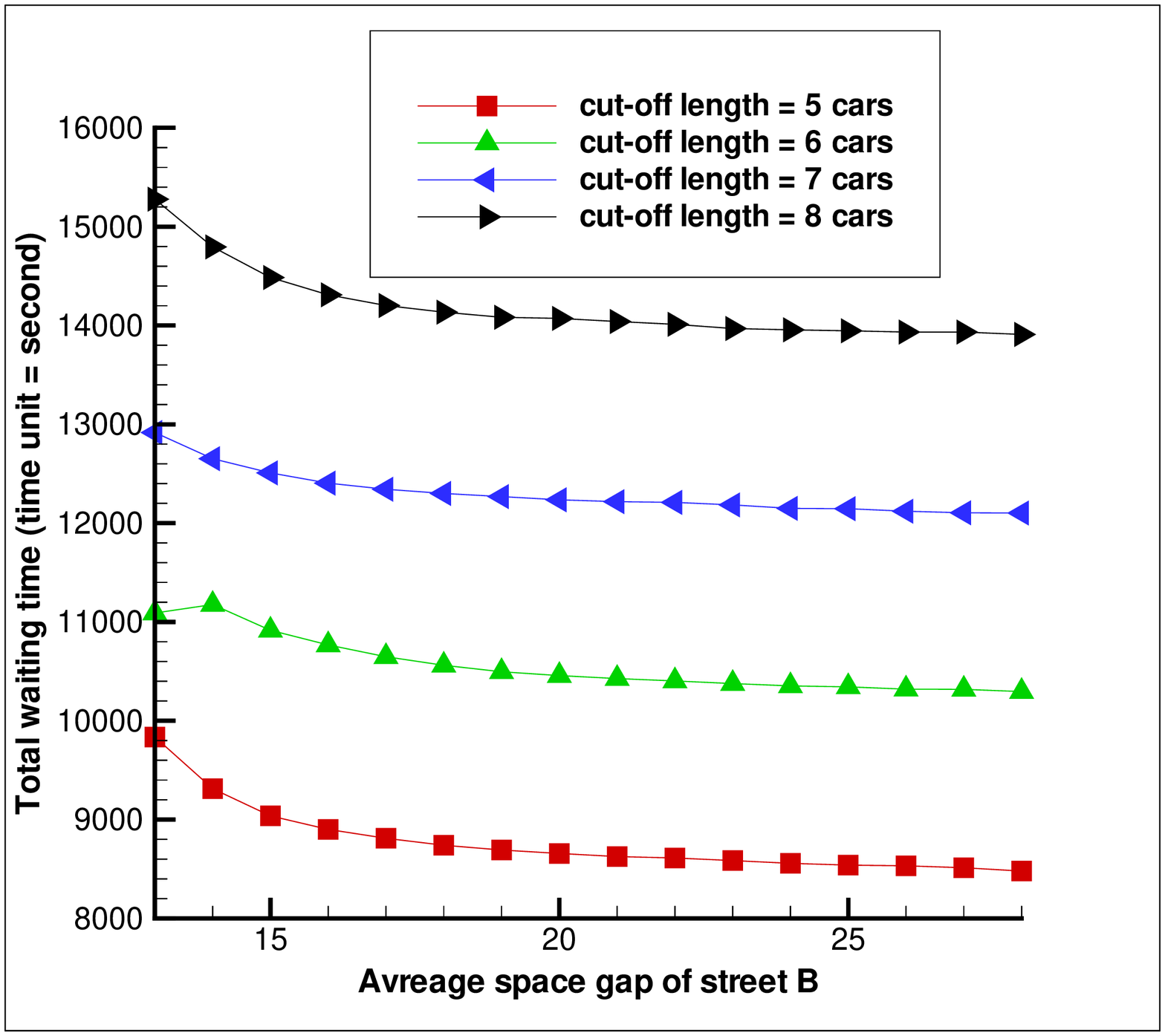}}
\end{figure}

\vspace{0.02 cm} {\small{Fig.~8: Total delay versus the inverse
traffic volume of street B for various cut-off lengths.
$\lambda_A$ is
fixed at 13.}}\\

We observe that decreasing the traffic volume in street B leads to
a decrease in total delay as expected. In the following figure,
the dependence of each street delay on the traffic volume of
street B is exhibited in details. One observes that the delay of
street B starts growing whereas that of street A decreases. This
shows that the algorithm does not always act optimally for
individual directions. In fact, the delay of street A, which is
more congested than street B, behaves decreasingly while the delay
of the less congested street, B, behaves increasingly.
Nevertheless, the algorithm acts in an optimal manner when taking
into account the whole intersection.

\begin{figure}\label{Fig9}
\epsfxsize=7.5truecm  \centerline {\epsfbox{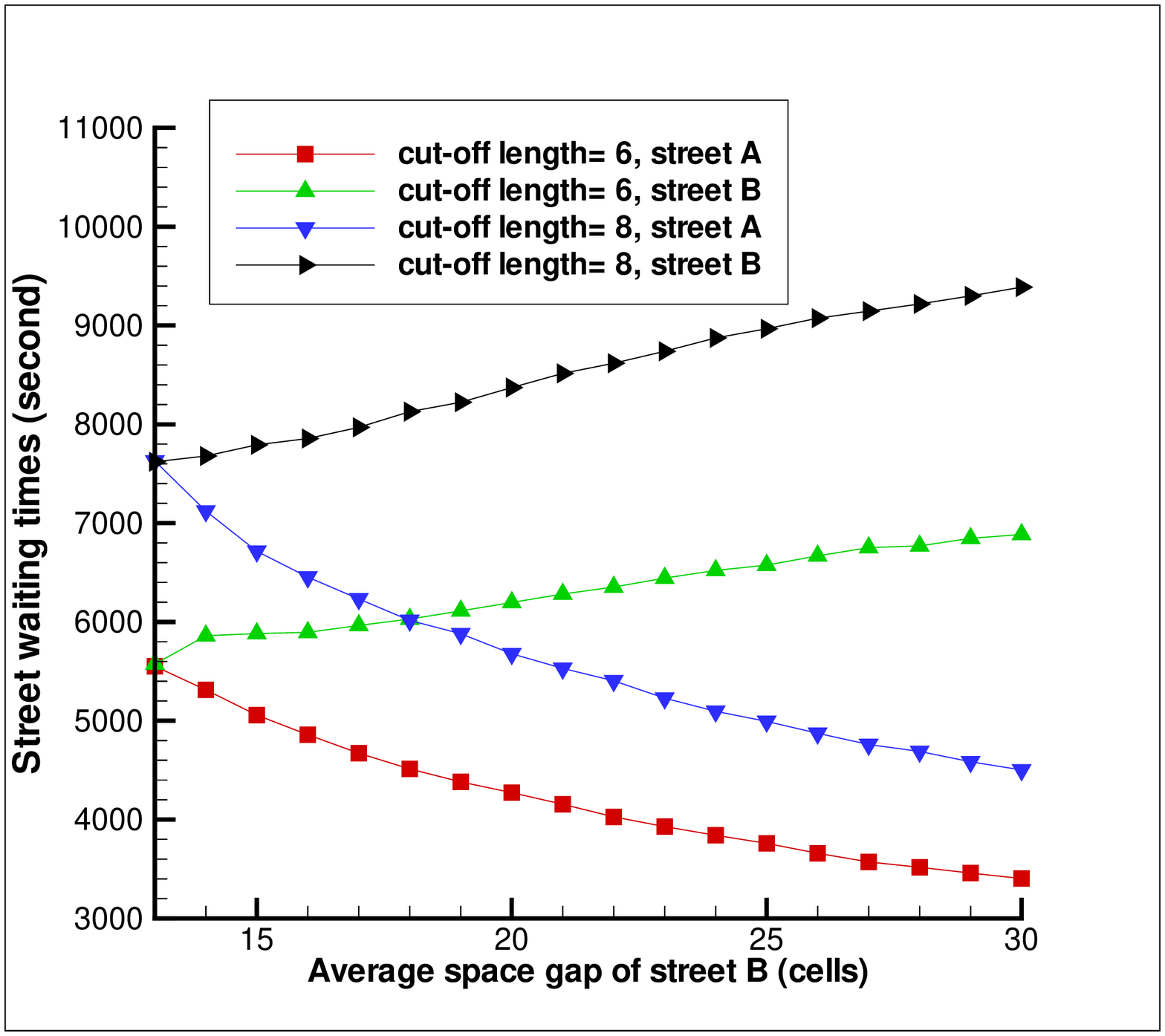}}

\end{figure}

\vspace{0.02 cm} {\small{Fig.~9: Delay of individual streets as a
function of inverse traffic volume of street B. Traffic volume of
street A is kept fix $\lambda_A=13$.}}\\

We have also examined the cases where cut-off lengths take
different values for streets with non-equal incoming flux.
Simulation results shows that optimal cut-off lengths should be
taken equal and as short as possible. For instance, in the case
$\lambda_A= 14$ and $\lambda_B= 24$ the minimised total delay when
$L_c^{(A)}=L_c^{(B)}=5$ is 8500 seconds.\\

{\bf scheme (2)}:\\

Let us now investigate the predictions of scheme (2). As
explained, in this method, once the global density of the moving
direction (in the vicinity of the intersection) falls below the
cut-off density, green phase is terminated irrespective of the
queue length in the red direction. The global density is obtained
by measuring the occupancy of $L_{\rho}$ cells before the crossing
point. The following graphs shows the one-hour delay behaviour of
individual streets as well as that of the entire intersection for
various controlling densities. Analogous to the predictions of
first scheme, here the algorithm increases the delay in the minor
road whereas the delay is decreased in the major road upon
decreasing the traffic volume in the minor road.

\begin{figure}\label{Fig10}
\epsfxsize=7.5truecm  \centerline {\epsfbox{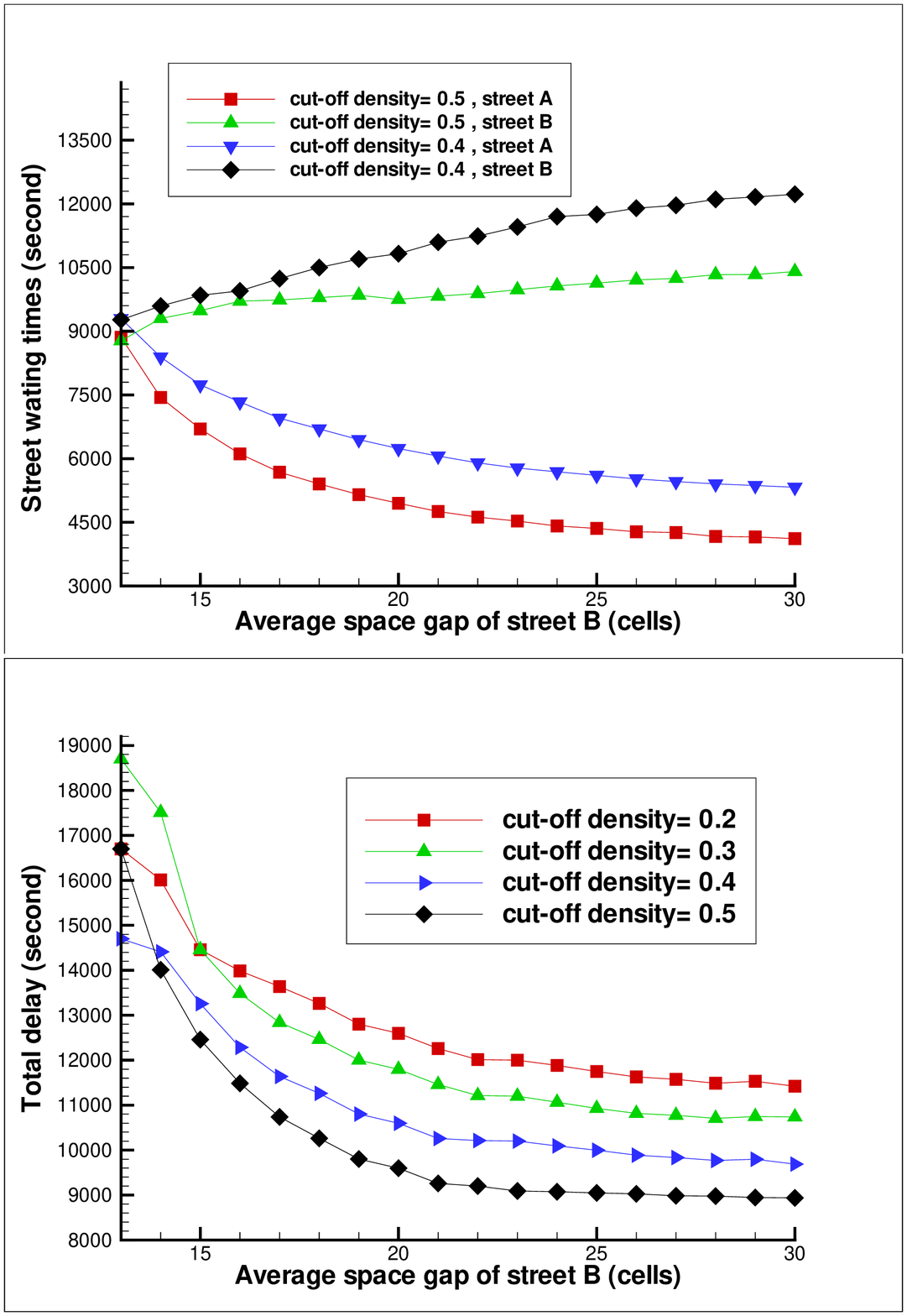}}

\end{figure}

\vspace{0.02 cm} {\small{Fig.~10: Top: delay of individual streets
as a function of inverse traffic volume of the minor street B.
Traffic volume of street A (major street) is kept fix
$\lambda_A=13$ cells. Bottom: total delay
versus the inverse traffic volume in the minor street for various controlling densities.}}\\

One observes that the delay predictions of the second scheme are
higher than those of the first scheme. We have also examined the
cases where cut-off densities of each street take non-equal
values. The following table exhibits the values of one-hour total
delay for various amounts of cut-off densities of each street.
Traffic
volumes correspond to $\lambda_A=13$ and $\lambda_B=18$ and $L_{\rho}$ is set to 10 cells.\\

\begin{tabular}{|c|c|c|c|c|c|}
\hline
$\rho_B \downarrow ~|~ \rho_A \rightarrow $ & 0.10 & 0.20 & 0.30 & 0.40 & $0.50$ \\
\hline
0.10  & 14800 & 17100 & 21000 & 20900 & $21100$  \\
\hline
0.20  & 13700 & 15400 & 15600 & 16700 & $16500$  \\
\hline
0.30  & 12900 & 13300 & 13700 & 14400 & $14600$  \\
\hline
0.40  & 11400 & 12100 & 13200 & 14200 & $14000$  \\
\hline
0.50  & 10800 & 11300 & 11800 & 12400 & $12900$  \\
\hline
0.60  & 10600 & 10800 & 10700 & 10900 & $11000$  \\
\hline
\end{tabular}
\vspace{.5cm}

We observe that optimal flow corresponds to set $\rho_A= 0.1$ and
$\rho_B= 0.6$. In contrast to scheme (1) where optimal cut-off
lengths are equal to each other, here we see that in asymmetric
traffic volume, the optimal cut-off densities are non-equal. We
note that other values of cut-off densities give rise to
unrealistic green time plans and hence are excluded from the
table. It would be useful to compare the minimal delay with the
predicted one in the scheme (1). For the traffic volumes
corresponding to $\lambda_A= 13$ and $\lambda_B= 18$, scheme (2)
gives the minimal delay at 10600 seconds whereas the scheme (1)
predicts the value 8700 seconds for $L_c=5$. We have also obtained
the total delay table for another asymmetric traffic volume where
in-flow rates are $\lambda_A= 13$ and $\lambda_B= 25$. For this
choice of traffic volumes, the optimal cut-off densities are
$\rho_A= 0.10$ and $\rho_B= 0.40$ which are relatively close to
the preceding optimal values. The values of minimal delays are
9300 and 8500 seconds in scheme (2) and (1) respectively. We have
investigated a variety of traffic states corresponding to
different in-flow rates. The results show that scheme (1) gives
better optimal states than scheme (2). Before discussing scheme
(3), we would like to discuss the role of parameter $L_{\rho}$. In
fact, the measurement of global density depends on the length
$L_{\rho}$. This length itself could be regarded as a control
parameter. We have carried out simulations for various values of
$L_{\rho}$. Our results shows that the $L_{\rho}=10$ corresponds
to the best choice. For instance the result obtained for the same
traffic state investigated in the above table and $L_{\rho}=12$
gives the optimal delay as 11000 seconds which is higher than the
value predicted by $L_{\rho}=10$.\\

{\bf scheme (3)}:\\

Finally we discuss the third scheme. In this scheme both cut-off
densities and cut-off lengths are taking into account. The
following graphs shows the one-hour delay behaviour of individual
streets as well as that of the entire intersection for various
controlling densities. Analogous to the predictions of first and
the second schemes, here the algorithm increases the delay in the
minor road whereas the delay is decreased in the major road upon
decreasing the traffic volume in the minor road. We have taken the
cut-off parameters equal for both streets i.e., $L_c^{A}=L_c^{B}$
and $\rho_c^{A}=\rho_c^{B}$.

\begin{figure}\label{Fig11}
\epsfxsize=7.5truecm  \centerline {\epsfbox{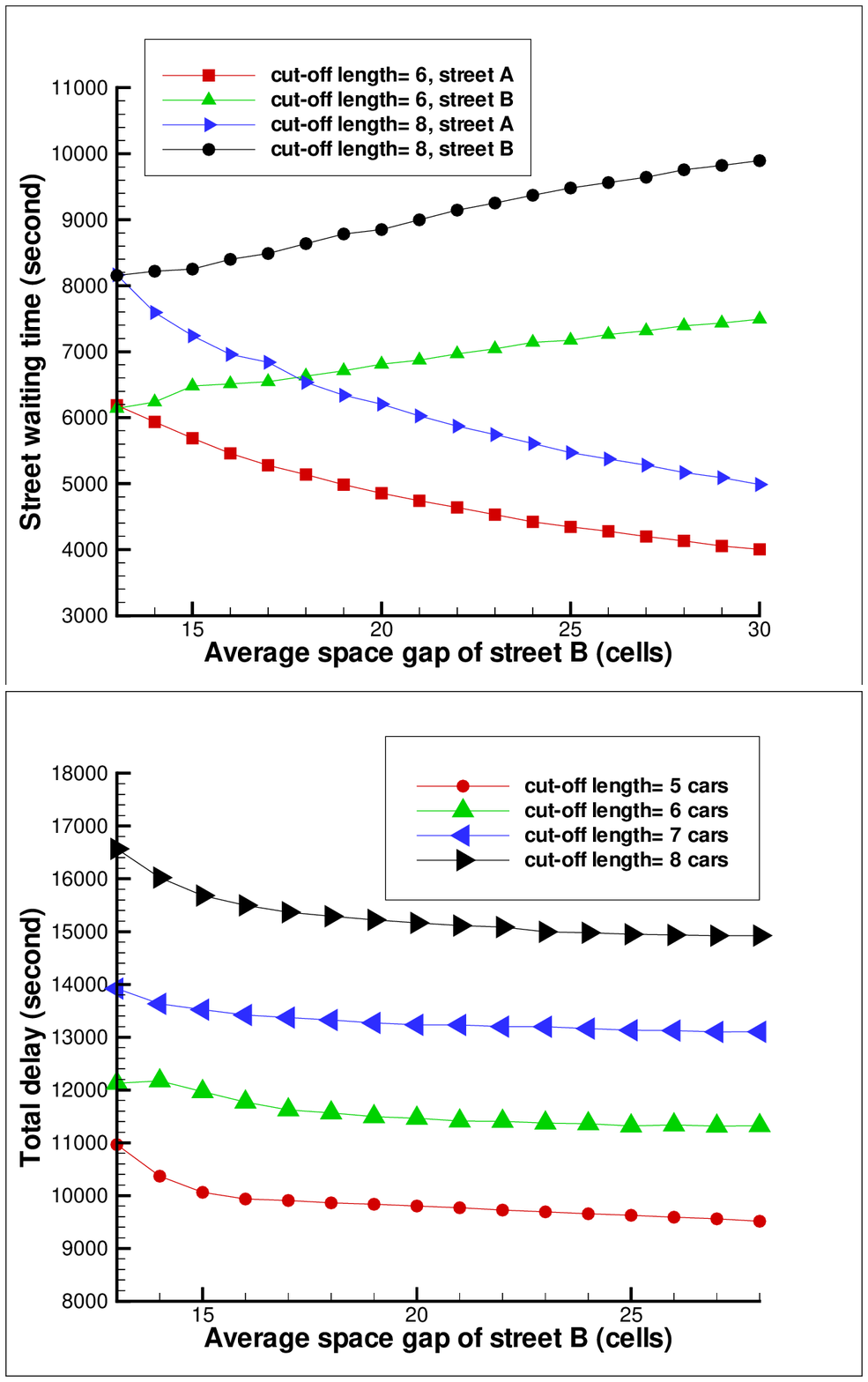}}

\end{figure}

\vspace{0.02 cm} {\small{Fig.~11: Top: delay of individual streets
as a function of traffic volume of the minor street B. Traffic
volume of street A (major street) is kept fix $\lambda_A=13$
cells. Bottom: total delay versus the traffic volume in the minor
street for various controlling densities. Cut-off densities are
taken 0.5 for both streets.}}\\


Another investigation considers the effect of cut-off densities
for fixed values of cut-off lengths. The following table exhibits
the delay when $\lambda_A= 13$ and $\lambda_B= 18$. $L_c=5$ while
$\rho_c$ is varied.\\

\begin{tabular}{|c|c|c|c|c|c|}
\hline
$\rho_c^{A} \downarrow ~ |~ \rho_c^{B} \rightarrow $ & 0.10
& 0.20 & 0.30 & 0.40 & 0.50
\\ \hline    0.10 & 12560 & 11600 & 11300 & 11000 & 10800
\\ \hline    0.20 & 12400 & 11370 & 10950 & 10600 & 10500
\\ \hline    0.30 & 12250 & 11200 & 10560 & 10300 & 9960
\\ \hline    0.40 & 12160 & 11150 & 10470 & 10200 & 9980
\\ \hline    0.50 & 11950 & 10780 & 10350 & 9970  & 9900
\\ \hline
\end{tabular}
\vspace{.5cm}

The table above shows that optimal traffic flow is obtained in the
high-high region of $(\rho_c^{A},\rho_c^{B})$ sub-space of control
parameters. The minimised delay is 9900 seconds. For the same
in-flow rates, the values of minimal delays predicted by the
schemes (2) and (1) are 10600 and 8700 seconds respectively. This
comparison shows that the third scheme acts better than scheme (2)
but is less efficient than the first scheme. Further simulation
results carried out for other traffic states demonstrates that for
$\lambda_B \leq 19$ ($\lambda_A$ fixed) scheme (3) acts better
than scheme (2) but is less efficient than scheme (2) for higher
in-flow rate of the minor street. Nevertheless, scheme (1) gives
the least delay for all asymmetric states.

\section{Summary and Concluding Remarks}

In this paper we have investigated the optimisation of vehicular
traffic flow at an isolated signalised intersection in the
framework of cellular automata. Marginal urban areas are places
where single intersections are frequently designed and operated.
Investigations on single junctions can be of practical relevance
for various applications in city traffic. Basically there are two
methods of signalisation of the traffic lights: fixed-time or
adaptive. In traffic adaptive strategies, it has always been a
subject of argument whether to control an intersection under a
centralized or decentralized scheme. In special circumstances,
decentralized local adaptive strategies operate more effectively
than globally adaptive strategies
\cite{book,robertson,porche,huberman,scats1,scats2} and often show
a very good performance. For this purpose, we have developed and
analysed prescriptions for the traffic light signalisation at
single intersections. Our simulation results confirm that in
fixed-time scheme, green times should be distributed proportional
to the traffic in-flow rates. We have simulated and analysed three
traffic responsive signalisation schemes. The signalisation
mechanisms are based on the concepts of cut-off queue lengths and
densities. Two major conclusion can be made from our simulation
results. First, the traffic adaptive scheme act more optimally
than the fixed-time. Secondly, the best traffic adaptive scheme is
the one in which the flow in the green direction is terminated as
soon as the queue length in the opposite direction exceeds the
cut-off value $L_c$. In this method, the traffic states in the
green direction are not taken into account. It should be noted
that the results in this paper have been obtained under a modified
version of Nagel-Schreckenberg model. Whether these result are
robust against more realistic vehicle movement models needs more
exploration and is the subject of our current investigation.
Throughout the paper we have used the Poisson statistics for the
headway of approaching cars. However the realistic in-flow traffic
would certainly deviate from this statistics. Investigations on
the impact of the entrance statistics on the above results will
shed more lights onto the problem.

\section{ Acknowledgments}

 We would like to express our gratitude to Nima Hamedani Radja for his fruitful and
 enlightening discussions. Special thanks go to Richard Sorfleet for reading the manuscript.
This paper is dedicated to Prof. Khajepour; co-founder of
Institute of Basic Studies in Advanced Sciences (IASBS); in the
occasion of his 60th birthday.

\bibliographystyle{unsrt}

\end{multicols}

\end{document}